\documentclass[aps,prb,twocolumn,amsfonts,amssymb,showpacs,eqsecnum]{revtex4}
\usepackage{amsmath}
\usepackage{amssymb}
\usepackage{graphicx}

\begin{document}

\title{The kinetics of escaping of Brownian particles
from a potential well for different space dimensionality. The effect
of external force.}
\author{A. I. Shushin}
\affiliation{Institute of Chemical Physics, Russian Academy of Sciences, 117977, GSP-1,
Kosygin str. 4, Moscow, Russia}

\begin{abstract}
The kinetics of two (2D) and three (3D) dimensional
diffusion-assisted escaping of Brownian particles from a potential
well in the presence of an external force is analyzed in detail. The
kinetics is studied within the two-state model (TSM) proposed for
processes in the absence of external force. The generalized variant
of this model, taking into account the force effect, is proposed
which is shown to be quite accurate for some shapes of the well both
for 2D and 3D processes. Within the generalized TSM simple
expressions for the well depopulation kinetics and, in particular,
for the escape rate are obtained. The effect of the force ($F$) is
shown to manifest itself in the escape rate dependence on the only
parameter $\varphi = Fa/(2k_b T)$, where $a$ is the Onsager radius
of the attractive part of the well $U(r)$, defined by the relation
$|U(a)| \approx k_b T$. The limiting behavior of this dependence in
the cases of weak and strong force is studied in detail both in 2D
and 3D processes. Some applications of obtained results to the
analysis of experiments are briefly discussed.

\end{abstract}

\maketitle

\bigskip

\section{Introduction}

The effect of external force on mechanisms and kinetic properties of
condensed phase diffusion-assisted reaction processes is considered
in a large number of works both experimentally and
theoretically.\cite{Ca,Ri,Yak,Mat} The active interest of scientists
to this phenomenon results from its great practical importance.

One of the most important systems, in which the external force
effect is investigated very thoroughly, is recombining geminate ion
pairs, undergoing relative diffusion in the external electric
field.\cite{Ri,Yak,Noo,Noo1} Most of theoretical studies analyze the
kinetics of the recombination process within the simplest model,
which reduces the problem to solving the Smoluchowski equation for
probability distribution function (PDF) of particles diffusing in a
pure Coulomb potential (with an external force) and reacting with
the rate highly localized at short
distances.\cite{Ri,Street1,Que,Woj} Even in this most simple
formulation the problem can, in general, be solved only numerically,
though detailed analytical analysis of some simple variant of the
problem have also been made,\cite{Ri,Street1} for example, within
the prescribed diffusion approximation.\cite{Mo}

Recent advances in time resolved investigations of charge transfer
and escaping processes in fast geminate reactions\cite{Chen} and, in
particular, geminate recombination of ion pairs in non-polar and
moderately polar
liquids\cite{Chen,Bart1,Bart2,Shkrob,Crow,Zhou1,Zhou2,Zhou3} inspire
further development of theoretical methods of the analysis of the
considered problem. The main challenge of the theoretical studies
consists in the correct description of the manifestation of specific
features of the interparticle interaction (in real liquids) in the
reaction kinetics in a tractable form simple enough to be suitable
for applications.

In the majority of above mentioned theoretical works no specific
features of the form of the interaction potential for the probe
(Brownian) particles at short distances (of order of molecular size)
have been taken into account. In the condensed phase, however, the
distance dependence of the potential at short interparticle
distances $r$ can be strongly modified by interaction of particles
under study with those of the medium.\cite{Ri} This modified
interaction is usually characterized by the so called mean force
potential (MFP), which in a physically reasonable form incorporates
the medium effect and, in particular, discreteness of the medium at
short distances. The interaction with the medium particles is known
to result in the wavy behavior of the MFP at short distances.
Moreover, in some systems the medium effect results in the well-type
shape of the MFP at short distances (see Fig. 1) with the a markedly
high barrier at distances $r$ of order of the distance of closest
approach $d$. This effect is found, for example, in the case of ion
pairs in polar liquids.\cite{Ri,Pat,Nin}

\begin{figure}
\setlength{\unitlength}{1cm}
\includegraphics[height=6cm,width=8.5cm]{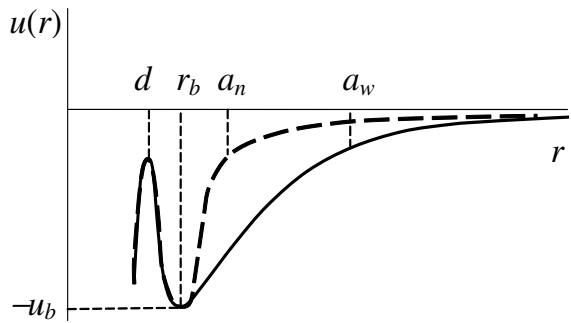}
\caption{The picture of the interaction potential $u(r)$ for two
models of its shape: narrow (dashed) and wide (full) well; $a_n$ and
$a_w$ are the Onsager radii for these models [defined by $u(a_{\nu})
= 1, \, (\nu = n, w)$], $d$ is the distances of closes approach, and
$r_{b}$ is the coordinate of the bottom.}
\end{figure}

Concerning the applicability of well-type approximations for the
real MFPs, it is also worth mentioning the additional reason: from
mathematical and kinetic points of view any attractive potential can
be considered as well-shaped in the absence of (or low) reactivity
of particles at $r \sim d$. The only difference of this type of
wells from those shown in Fig. 1 is in their urge-like shape at $r
\sim d$.

The well-type shape of the MFP (with the reaction barrier at $r \sim
d$) results in the formation of the quasiequilibrium state within
the well, which can be considered as a cage. In the absence of
external force the kinetics of diffusion-assisted depopulation of
the initially populated cage state is analyzed in detail in a number
of papers.\cite{shu1,shu2,shu3,shu4} In the limit of deep well depth
the problem is shown to be accurately described with the two-state
model (TSM), i.e. the model of two kinetically coupled states: the
quasiequilibrium localized state within the well and the free
diffusion state outside the well.\cite{shu3}

The TSM enables one to obtain the well depopulation kinetics in a
relatively simple analytical form. This kinetics, determined by the
monomolecular reactive passing over the barrier at $r\sim d$ and
escaping from well (cage), appears to be non-exponential, in
general. In the limit of deep well, however, the deviation from the
exponential kinetics is shown to be fairly small.\cite{shu2,shu3}

In this paper we generalize the TSM to describe the effect of the
external force on the well depopulation kinetics. The effect is
assumed to result only from the force induced change of the rate of
escaping from the well and will concentrate on the discussion of the
manifestation of the force in the escaping kinetics.

Within the generalized TSM we derive simple formulas for diffusion
assisted well depopulation (or escaping) kinetics in two (2D) three
(3D) dimensional spaces for different models of the well shape.
Moreover, taking into account that in the limit of deep well (which
is of main interest of the discussion) the kinetics is close to
exponential, special attention is paid to the analysis of the escape
rate. The analysis shows that the specific features of the force
effect on the escape rate depends on the well shape. This effect can
be characterized by the rate dependence on the only parameter. In
the cases of weak and strong force the limiting analytical
expressions for this dependence are obtained and briefly discussed.

Some possible applications of obtained results are also discussed.
It is shown that the kinetic formulas are fairly useful for the
analysis of different experiments. As examples we considered recent
time resolved experiments on liquid phase chemical reactions
\cite{Chen,Bart1,Bart2,Shkrob}, transient photocurrents arising from
dissociation of exciplexes in solutions
\cite{Zhou1,Zhou2,Zhou3,Lukin}, and the kinetics of colloidal
particle trapping into and escaping from the optical force induced
potential well (tweezers) \cite{Ashkin,Grier}.

\section{Formulation of the problem}

We consider the dynamics of the Brownian particle undergoing (2D or
3D) diffusive motion in a spherically symmetric MFP well $U({\bf r})
\equiv U(r)$ centered at $r = 0$, where ${\bf r}$ is the vector of
position of the particle. The dependence of $U(r)$ on the distance
$r = |{\bf r}|$ is schematically shown in Fig.1. The MFP will be
characterized by three parameters: the distance $a$ (called
hereafter the Onsager radius), at which $U(a) \approx k_B T$, which
is defined as:
\begin{equation} \label{form00}
a = \left(\int_{r_b}^{\infty} \! dr \, r^{-2}e^{u(r)}\right)^{-1}
\end{equation}
[in Fig. 1 $a_n$  and $a_w$ denote the Onsager radii, corresponding
to two models of the well shape, discussed below (in Sec. V)], the
distance of closest approach $d$, and the radius $r_b$ of the bottom
of the well whose energy is $U(r_b)= -U_b = -(k_B T) u_b$. At $r
\sim d$ the MFP $U(r)$ is assumed to be of the shape of a barrier,
diffusive passing over which models the reaction within the well.

The main purpose of the work is to analyze the kinetics of
diffusion-assisted escaping from the well in the presence of the
external force $-{\bf F}$, i.e. escaping from the well of the
potential $U_{\bf f}=U(r)+({\bf F}\cdot{\bf r})$. For definiteness
the force is assumed to be directed along the axis $z$: ${\bf F}=
(0,0,F)$. The analysis can conveniently be made in spherical
coordinates in which ${\bf r}= (r\sin \theta \cos\phi, r \sin \theta
\sin\phi, r \cos\theta)$.

The diffusive space-time evolution of the Brownian particle is
described by the PDF $\rho ({\bf r},t)\equiv \rho
(r,\theta,\phi|t)$. In general, in spherical coordinates the PDF
depends on all three variables $r$, $\theta$, and $\phi$. However,
in the considered case of isotropic diffusion and isotropic initial
condition $\rho ({\bf r},t=0) = \rho_i (r)$ the PDF $\rho ({\bf
r},t)$ is independent of the azimuthal angle $\phi$ so that $\rho
({\bf r},t) \equiv \rho (r,\theta|t)$. In our analysis we will
assume that particles are created within the well at initial
distance $r_i \sim r_b$:
\begin{equation} \label{form0}
\rho ({\bf r},t=0)= \rho_i ({\bf r}) = N_i\delta (r - r_i),
\end{equation}
where $N_i = (2\pi r_i)^{-1}$ and $N_i = (4\pi r_i^2)^{-1}$ for 2D
and 3D processes, respectively.

The PDF $\rho (r,\theta|t)$ satisfies the Smoluchowski equation
\begin{equation} \label{form1}
\dot \rho =  \nabla_{\bf r}[ D (r) (\nabla_{\bf r} \rho +
\rho\nabla_{\bf r}u_{\bf f}) ],
\end{equation}
where $\nabla_{\bf r} $ is the gradient operator,
\begin{equation} \label{form2}
u_{\bf f}({\bf r}) =u(r)+({\bf f}\cdot{\bf r}),
\end{equation}
with $u(r)= U(r)/(k_B T)$ and ${\bf f}= {\bf F}/(k_B T)$, is the
dimensionless MFP, and $D $ is the diffusion coefficient for the
particle, which in our analysis is assumed to be independent of $r$
[though some possible effects of $D(r)$-dependence can also be
studied].

Note that the reaction kinetics for pairs of interacting Brownian
particles, say $a$ and $b$, is described by the equation similar to
eq. (\ref{form1}) with ${\bf r} = {\bf r}_a-{\bf r}_b$ and
parameters expressed in terms of those for separate
particles.\cite{Ri,Noo1}

In the absence of force the kinetics of escaping from the
spherically symmetric short range potential well $u (r)$ is analyzed
earlier.\cite{shu1,shu2,shu3,shu4} Here we extend the approach
applied in these works to describe the effect of external force.
This approach is based on the approximate solution of eq.
(\ref{form1}) in the limit of deep well, in which this equation can
be solved analytically by expansion in the small parameter $\tau_c
/\tau_e \ll 1$, where
\begin{equation} \label{form3}
\tau_c \sim (a-d)^2/D \;\;\mbox{and}\;\;\tau_e \sim \tau_c e^{u_b}
\end{equation}
are the time of equilibration within the well and the time of
escaping from the well, respectively.\cite{shu2,shu3}

Analysis of this solution shows\cite{shu3} that in the lowest order
in the parameter $\tau_c /\tau_e$ the Smoluchowski approximation
(\ref{form1}) is equivalent to the model of two kinetically coupled
states: the state within the well and the state of free diffusion
outside the well.

\section{Two-state model. 3D processes.}

Originally, the TSM was proposed to treat the kinetics of
diffusion-assiisted escaping from the well in the case ${\bf F} =
0$, in which the PDF $\rho ({\bf r},t)$ depends only the distance $r
= |{\bf r}|$. In the presence of force [${\bf f} = {\bf F}/(k_B
T)\neq 0$], however, the potential $u_{\bf f}(r)=u(r)+({\bf
f}\cdot{\bf r})$ in eq. (\ref{form1}), is anisotropic, which results
in the dependence of the solution $\rho ({\bf r},t)$ on particle
polar angle $\theta$: $\rho ({\bf r},t) \equiv \rho (r,\theta|t)$.

It is important to note that in the case ${\bf f} \neq 0$ the TSM is
also valid for a variety of shapes of the potential well $u (r)$
though some additional analysis of the corresponding validity
criteria are certainly required.

\subsection{Kinetic equations}

In accordance with rigorous consideration\cite{shu2,shu3}, in the
lowest order in $\tau_c /\tau_e \sim e^{-u_a} \ll 1$ the escaping
kinetics can be described within the model of two kinetically
coupled states: highly localized state within the well and free
diffusion state outside the well. The evolution of the state within
the well ($d < r < a$) is determined by the well population
\begin{equation} \label{nof2a}
n(\theta|t) = 4\pi \int_d^{a} \! dr \,r^2 \rho (r,\theta|t),
\end{equation}
while the evolution of the state outside the well ($r > a$) is
governed by the PDF $c(r,\theta|t)$.  Within the TSM the effect of
the force ${\bf f}$ shows itself in the dependence of kinetic
parameters on $\theta$. The form of this dependence is determined by
the particular variant of the model.

In general, TSM kinetic equations, describing evolution of PDFs
$n(\theta|t)$ and $c(r,\theta|t)$ in the presence of an external
force, can be written as:\cite{shu2,shu3}
\begin{subequations} \label{prf7}
\begin{eqnarray}
\dot n &=& S_a K_{+}(\theta) c(a|t) + [\hat L_{c}-( K_{-}(\theta) +
w_{r})]n, \qquad
 \label{prf7a}\\
\dot c &=& \hat L_{f} c + [S_{a}^{-1}K_{-}(\theta) n - K_{+}(\theta)
c] \delta (r-a),\qquad \label{prf7b}
\end{eqnarray}
\end{subequations}
where $S_a = 4\pi a^2$,
\begin{equation} \label{prf8}
\hat L_{f} = D\nabla_{\bf r}(\nabla_{\bf r}  + {\bf f})
\end{equation}
is the operator, describing 3D-diffusion outside the well, and $\hat
L_c$ is the Smoluchowski operator in $\{\theta\}$-space which
controls orientational relaxation of the PDF in the well.

The essential difference of anisotropic equations from isotropic
ones consists in the orientation dependence of rates,
$K_{+}(\theta)$ and $K_{-}(\theta)$. In the considered limit $\tau_c
/\tau_e \ll 1$ we will assume the transition rates $K_{\pm}$ to
satisfy the relations:\cite{shu3}
\begin{equation} \label{nof7c}
K_{\pm} \rightarrow \infty \;\;\mbox{and}\;\;
K_{-}(\theta)/K_{+}(\theta) = K_e(\theta).
\end{equation}
Therefore in this limit $\theta$-dependence of rates
$K_{\pm}(\theta)$ show itself in that of the the equilibrium
constant: $K_e(\theta)$. The form of the function $K_e(\theta)$ is
determined by the shape of the well. Some model well shapes and
corresponding $K_e(\theta)$ dependences, as well as applicability of
the corresponding TSMs, are discussed below.

Equations (\ref{prf7}) should be solved with boundary conditions
\begin{equation} \label{prf9}
(\nabla_r + f\cos\theta) c|_{r=a}=0
\;\;\mbox{and}\;\;c|_{r\to\infty} \to 0,
\end{equation}
first of which describes reflection of particles (diffusing in the
state outside the well) at $r = a$. The initial condition is assumed
to be isotropic and given by eq. (\ref{form0}).

In what follows it will be convenient to represent functions
$n(\theta|t)$ and $c(r,\theta|t)$ in the form of vectors $|{\bf n}
(t)\rangle$ and $|{\bf c} (r,t)\rangle$, whose components are
obtained by expansion of these functions in the orthonornmal basis
of properly normalized Legendre polynomials (spherical
functions)\cite{Lan}
\begin{equation}
|l\rangle = (l + \mbox{$\frac{1}{2}$})P_l (\cos\theta), \;\,\langle
l| = \int_0^{\pi}\!\!d\theta  \sin\theta P_l
 (\cos\theta) \dots \label{prf1a}
\end{equation}
with $l=0,1,\dots:$
\begin{equation} \label{prf2}
|{\bf n}\rangle = \sum_{l=0}^{\infty} n_l |l\rangle  \;\; \mbox{and}
\;\; |{\bf c}\rangle  = \sum_{l=0}^{\infty}  c_l |l\rangle,
\end{equation}
where for any vector $|\chi (\theta)\rangle,\; (\chi = n, c),$ its
components $\chi_l$ are defined by
\begin{equation} \label{prf3}
\chi_l  = \langle l | \chi\rangle = \int_{0}^{\pi} \!\! d\theta\,
\sin\theta\, P_l(\cos\theta)\chi(\theta).
\end{equation}

In term of this vector representation the initial condition can
conveniently be written in the form, explicitly displaying its
independence of orientation:
\begin{equation} \label{prf10}
|\rho_i \rangle =(2\pi r_i^2)^{-1}|0\rangle \delta (r - r_i).
\end{equation}

As for the initial condition, it is worth noting, in addition, that
in the most realistic limit of orientational relaxation within the
well much faster than the escaping from the well the escaping
kinetics is insensitive to the orientational dependence of the
initial condition.

\subsection{Escaping kinetics}

Equations (\ref{prf7}) can be solved by the method applied in the
absence of force but with the use of expansion of $n(\theta|t)$ and
$c(r,\theta|t)$ in spherical functions $|Y_l\rangle$, i.e. vector
representation $|{\bf n}(t)\rangle$ and $|{\bf c}(r|t)\rangle$ [see
eq. (\ref{prf2})]. The solution yields for the Laplace transform
\begin{equation} \label{narw1}
|\widetilde{\bf n}(\epsilon)\rangle =  \int_0^{\infty}\!dt
e^{-\epsilon t}|{\bf n}(t)\rangle:
\end{equation}
\begin{equation} \label{narw2}
|\widetilde{\bf n}(\epsilon)\rangle =  \big[\epsilon + w_r - \hat
L_{c} + \hat W_e (\epsilon)\big]^{-1}|{\bf n}_i \rangle.
\end{equation}
In this expression
\begin{equation} \label{nof8a}
w_r = (D/Z_w) \left(\int_{r\sim d} \! dr \,
r^{-2}e^{u(r)}\right)^{-1} .
\end{equation}
is the rate of first order reaction in the well in which
\begin{equation}\label{nof7}
Z_w = \int_{d<r<a}dr \, r^2 e^{-u(r)}
\end{equation}
is the partition function for the well $u(r)=U(r)/(k_B T)$ [in the
absence of force ($f = 0$)],\cite{shu2,shu3}
\begin{equation} \label{narw4}
\hat W_e(\epsilon) = \hat G^{-1}(a,a|\epsilon)\hat K_e
\end{equation}
is the operator of the generalized  $\epsilon$-dependent escape rate
(see below), in which the equilibrium constant $\hat K_e$ is the
operator, indicating its dependence on the angle $\theta$, and $\hat
G(a,a|\epsilon)$
is the evolution operator for diffusive motion outside the
well (evaluated at $r = r_i = a$):
\begin{eqnarray} \label{narw5}
\hat G(a,a|\epsilon) &=& \langle a | (\epsilon - \hat
L_f)^{- 1} |a \rangle\nonumber\\
&=& e^{-\varphi\cos\theta}\langle a | (\epsilon - \hat \Lambda_f)^{-
1} |a \rangle e^{\varphi\cos\theta},\qquad
\end{eqnarray}
where $\varphi = fa/2$ and
\begin{equation} \label{narw6}
\hat \Lambda_f = D\big(\hat L_r + r^{-2}\hat
L_{\theta}-\mbox{$\frac{1}{4}$}f^2\big),
\end{equation}
is the auxiliary operator in which
\begin{equation} \label{narw7}
\hat L_{\theta} =
\frac{1}{\sin\theta}\nabla_{\theta}\big(\sin\theta\,\nabla_{\theta}\big)
= -\sum_{l=0}^{\infty}l(l+1)|l \rangle \langle l|.
\end{equation}
and
\begin{equation} \label{nof5}
\hat L_r = D r^{-2}\nabla_r(r^2 \nabla_r)
\end{equation}
are the operators of free orientational and radial diffusion,
respectively.

Noteworthy is that for the initial condition (\ref{prf10}) $|{\bf
n}_i \rangle = (1/2\pi) |0\rangle $.

In what follows we will restrict ourselves to the analysis of the
well depopulation kinetics $\bar n(t)$, whose Laplace transform is
given by
\begin{eqnarray} \label{narw8a}
\widetilde{\bar n}(\epsilon) &=& 2\pi\!\int_0^{\pi}\!\!d\theta\,
\sin\theta \widetilde{n}(\theta,\epsilon) \equiv
2\pi\langle 0|\widetilde{\bf n}(\epsilon)\rangle\nonumber\\
& = &\langle 0|\big[\epsilon + w_r - \hat L_{c} + \hat W_e
(\epsilon)]^{-1}\big|0 \rangle.
\end{eqnarray}

\subsection{General formulas}

Formulas (\ref{narw5}-(\ref{narw7}) allow us to evaluate the
operator $\hat G^{-1}(a,a|\epsilon)$ in analytical form and,
therefore, analyze the behavior of $\widetilde{n}_0(\epsilon)$
relatively easily.

In the evaluation it is worth taking into account the useful
relation which simplifies the differential operator in the radial
space:
\begin{equation} \label{narw9}
\langle a | (\epsilon - \hat \Lambda_f)^{- 1} |a \rangle = \langle a
| (\epsilon - \hat \lambda_f)^{- 1} |a \rangle,
\end{equation}
where
\begin{equation} \label{narw11}
\hat \lambda_f = D(\nabla_r^2 + r^{-2}\hat
L_{\theta}-\mbox{$\frac{1}{4}$}f^2).
\end{equation}

The evolution operator $\langle a | (\epsilon - \hat \lambda_f)^{-
1} |a \rangle$ can be obtained in analytical form\cite{shu5} with
the use of two linearly independent operator solutions $\hat
\psi_{-} (r)$ and $\hat \psi_{+} (r)$ of equation
\begin{equation} \label{narw12}
(\epsilon - \hat \lambda_f)\hat {\bf \psi}_{\pm} = 0
\end{equation}
in which the operator $\hat L_{\theta}$ is treated as a parameter.
These solutions satisfy two boundary conditions corresponding to
those given in eq. (\ref{prf9}) [after change of variable $c(r) =
e^{-(fr\cos\theta)/2}\,\psi (r)$]
\begin{equation} \label{narw14}
(\nabla_r + \mbox{$\frac{1}{2}$}f\hat \omega) \hat \psi_{-}|_{r=a}=0
\;\;\mbox{and}\;\; \hat \psi_{+}|_{r\to\infty} \to 0,
\end{equation}
where
\begin{equation} \label{narw15}
\hat \omega = \sum_{l,l^{\prime}=0}^{\infty} |l\rangle \langle l
|\cos\theta |l^{\prime}\rangle\langle l^{\prime} |
\end{equation}
is the matrix representation of the function $\cos\theta$. The
matrix elements $\langle l|\cos\theta |l^{\prime}\rangle = \langle
l|P_1(\cos\theta) |l^{\prime}\rangle$ are evaluated
analytically\cite{Lan} though the corresponding formulas will not be
needed in our further analysis.

Both solutions $\hat \psi_{+} (r)$ and $\hat \psi_{-} (r)$ are
expressed in terms of matrices of Bessel functions
$K_{l+\frac{1}{2}}(x)$ and $I_{l+\frac{1}{2}}(x)$,\cite{Abr}
\begin{equation} 
\hat X (r) = \sqrt{r}\sum_{l=0}^{\infty}|l \rangle
X_{l+\frac{1}{2}}(\varphi_{\epsilon}r/a)\langle l|, \;\:
(X = I, K)\!:\!\!\!\label{narw17a}\\
\end{equation}
\begin{equation} \label{narw18}
\hat \psi_{+} (r) = \hat K (r),\;\;\hat \psi_{-} (r) = \hat I (r) +
\hat K (r)\hat\kappa,
\end{equation}
where $\varphi = fa/2$,
\begin{equation} \label{narw18a}
\varphi_{\epsilon} = \varphi\sqrt{1 + \epsilon/\epsilon_f}, \;\;\;
\epsilon_f = Df^2/4 = (D/a^2)\varphi^2,
\end{equation}
and $\hat \kappa$ is the matrix determined by the boundary condition
at $r =a$ [see eq. (\ref{narw14})]:
\begin{equation} \label{narw19}
\hat\kappa = [\nabla_r \hat K(r) -\hat q \hat K(r)]^{-1}[\hat q \hat
I(r)-\nabla_r \hat I(r)]|_{r=a},
\end{equation}
in which
\begin{equation} \label{narw20}
\hat q = a^{-1}(1 - \varphi \hat \omega)\equiv a^{-1}(1 - \varphi
\cos\theta).
\end{equation}

It is worth noting that the matrices $\hat K$ and $\hat I$ do not
commute with $\hat \omega$ and, therefore, the order of matrices in
the products of the matrices in expressions
(\ref{narw18})-(\ref{narw19}) is important. As a result of these
special commutation properties of the matrices, the matrix solutions
$\hat \psi_{+} (r)$ and $\hat \psi_{-} (r)$ do not commute either.

The representation of the evolution operator  $\langle r | (\epsilon
- \hat \lambda_f)^{- 1} |r_i \rangle$ in terms of non-commuting
solutions $\hat \psi_{+} (r)$ and $\hat \psi_{-} (r)$ is proposed
and thoroughly discussed in ref. [29]. This representation
generalizes the well known one for scalar solutions $\psi_{+} (r)$
and $\psi_{-} (r)$. In general, the proposed representation is
fairly cumbersome. In the particular case of solutions given by eq.
(\ref{narw17a}), however, it reduces to a more simple one:
\begin{equation} \label{narw21}
\langle r | (\epsilon - \hat \lambda_f)^{- 1} |r_i \rangle = \hat g
(r,r_i) + \hat K (r)\hat \kappa \hat K(r_i)W_{\hat I \hat K}^{-1}
\end{equation}
In this formula
\begin{equation} \label{narw22}
W_{\hat I \hat K} = D[\nabla_r \hat I(r) \hat K(r) - \nabla_r \hat
K(r) \hat I(r)] = D
\end{equation}
is the Wronskian of two solutions and
\begin{eqnarray} \label{narw21a}
\hat g (r,r_i) &=& [\hat K (r)\hat I(r_i)\theta(r-r_i) \nonumber \\
&& + \hat I(r)\hat K (r_i)\theta(r_i-r)]W_{\hat I \hat K}^{-1}
\end{eqnarray}
is the evolution operator for $\hat \kappa = 0$ in which
$\theta_H(x)$ is the Heaviside step function.

The validity of the expression (\ref{narw21}) can be verified by
direct substitution to equation inhomogeneous variant of eq.
(\ref{narw12}) with delta-function in the right hand side.

For the particular case $r=r_i=a$ formula (\ref{narw21}) yields
\begin{equation} \label{narw23}
\langle a | (\epsilon - \hat \lambda_f)^{- 1} |a \rangle =
D^{-1}[\hat q + \hat q_K(\epsilon)]^{-1},
\end{equation}
where
\begin{equation} \label{narw24}
\hat q_K(\epsilon) = -\nabla_r \hat K(r)/\hat K(r)|_{r=a} =
\sum_{l=0}^{\infty}|l\rangle q_{K_l}(\epsilon) \langle l|
\end{equation}
with
\begin{equation} \label{narw25}
q_{K_l}(\epsilon) = a^{-1}\big[l + \varphi_{\epsilon}
K_{l-\frac{1}{2}} (\varphi_{\epsilon})/K_{l+\frac{1}{2}}
(\varphi_{\epsilon})\big].
\end{equation}

Substituting the expression (\ref{narw23}) into eq. (\ref{narw9})
and then into eqs.(\ref{narw6}) and (\ref{narw4}) we obtain formula
\begin{equation} \label{narw26}
\hat G^{-1}(a,a|\epsilon) = D[\hat q + e^{-\varphi\cos\theta}\hat
q_K (\epsilon) e^{\varphi\cos\theta}].
\end{equation}

For our further analysis of the escaping kinetics $n_0 (t)$ we need
to specify of the operator $\hat L_c$ describing orientational
relaxation in the well. Naturally it should be of the
Smoluchowski-like form:
\begin{equation} \label{narw27}
\hat L_{c} = D_c(\sin\theta)^{-1}
\nabla_{\theta}[\sin\theta(\nabla_{\theta} + \nabla_{\theta}\bar
u)],
\end{equation}
where $D_c \sim D/r_b^2$ is the orientational diffusion coefficient
$\bar u (\theta)$ is the effective orientational potential which is
determined by the shape of the well (see below).


Moreover, in the considered limit of large well depth it is quite
natural to assume that orientational relaxation is much faster than
well depopulation.

\subsection{Fast orientational relaxation in the well}

The fast orientational relaxation limit implies that $D_c \gg
\tau_e^{-1}$.
This relation means that after some time $\sim \tau_c = D_c^{-1}$ of
orientational relaxation (of the initial population in the well) the
vector of well population $|{\bf n}(t)\rangle$ remains close to the
equilibrium one $|\Psi_e\rangle$ during the process:
\begin{equation} \label{narw28}
|{\bf n}(t)\rangle \approx \bar n(t) |\Psi_e\rangle,
\end{equation}
where
\begin{equation} \label{narw29}
|\Psi_e\rangle = Z_{\theta}^{-1}e^{-\bar{u}_b(\theta)},
\;\;\;\; Z_{\theta} =\int_{0}^{\pi}\!\!d\theta \,\sin\theta
e^{-\bar{u}_b(\theta)}.
\end{equation}
Note that  within bra-ket notation the adjoint vector $\langle
\psi_e |$ coincides with $\langle Y_0 |$ and is given by formula
\begin{equation} \label{narw29a}
\langle \Psi_e | = \langle 0 | = \int_0^{\pi}\!d\theta \, \sin\theta
\dots,
\end{equation}
which can be confirmed by the relation $\langle \Psi_e |\hat L_{c} =
0$ directly following from the definition of $\hat L_{c}$ [see eq.
(\ref{narw27})]. With the use of this formula one can easily find
that $|\psi_e \rangle$ satisfies the normalization condition
$\langle \Psi_e|\Psi_e \rangle = 1$.

In what follows we will restrict ourselves to the analysis of the
escaping kinetics just in this limit of fast orientational
relaxation.

For fast orientational relaxation the splitting $\delta L_c$ of
eigenvalues of the operator $\hat L_c$ ($\delta L_c \sim D_c$) is
much larger than $\|\hat W_e \| \sim w_e^0$. In such a case in the
lowest order in the parameter $w_e^0/D_c \ll 1$ we can significantly
simplify the general expression for $\widetilde{n}_0 (\epsilon)$
[eq. (\ref{narw8a})] and thus for the inverse average lifetime
${\bar w}_0 = {\bar \tau}_0^{-1} = \widetilde{n}_0^{-1} (0)$ as
follows:
\begin{equation} \label{narw30}
\widetilde{\bar n}(\epsilon)  = \big[\epsilon \!+\! w_r \! + \!
w_e(\epsilon)]^{-1} \;\:\mbox{and}\;\: \bar w_0 = w_r  + \bar w_e,
\end{equation}
where
\begin{equation} \label{narw30a}
w_e(\epsilon) = \langle \Psi_e|\hat W_e(\epsilon) |\Psi_e\rangle
\;\;\mbox{and}\;\; \bar w_e = w_e(0).
\end{equation}

Equation (\ref{narw30}) presents the main result of the work for the
kinetics of the well depopulation in the limit of fast orientational
relaxation.

The depopulation kinetics predicted by formulas (\ref{narw30}) and
(\ref{narw30a}) essentially depends on the shape of the potential
well which determines the orientational potential $\bar u (\theta)$
in the Smoluchowski-type operator $\hat L_c$ [see eq.
(\ref{narw27})] and, therefore, the equilibrium state
$|\Psi_e\rangle $. In our work we will consider two realistic models
of the well shape in which simple analytical expressions well
depopulation kinetics
can be obtained.

\subsection{Application of results}

In this section we will analyze the specific features of mean escape
rate $w_e$ for two variants of the well shape:

1) Narrow well shape (shown in Fig. 1 by dashed line with $a =
a_n$), for which $a - d \ll d$ and the time of equilibration within
the well $\tau_c \sim (a-d)^2/D \ll a^2/D \ll \tau_e$, where
$\tau_e$ is the time of escaping from the well [see eq.
(\ref{form3})];

2) The wide well shape (full line in Fig. 1 with $a = a_w$),
corresponding to a small distance of closest approach $d \ll a$ (in
which of the main interest is the region  $fd \ll 1$ while $fa
\gtrsim 1$). This shape is schematically shown in Fig. 1 by full
line with $a = a_w$.

The analysis will be made in the above-discussed limit of fast
orientational relaxation in the well with the use of eq.
(\ref{narw30}) for the well depopulation kinetics. In our study, in
addition to this formula we will also use the representation for $
e^{\pm\varphi \cos\theta}$ in terms of expansion in spherical
functions $P_{l}(\cos\theta)$:\cite{Abr}
\begin{equation} \label{narw31}
 e^{\pm\varphi\cos\theta} = \! \sqrt{\frac{2\pi}{\varphi}}\,\sum_{l=0}^{\infty}(\pm
 1)^{l} (l + \mbox{$\frac{1}{2}$}) I_{l+\frac{1}{2}}(\varphi)P_{l}(\cos\theta).
\end{equation}

\subsubsection{Narrow-well shape}

In the case of narrow well, when $a -d \ll d$, the well is of the
shape of attractive well layer near the distance of closest approach
$d$. In this limit within the wide region force strengths $f <
1/(a-d)$ we can neglect the effect of the force on the radial shape
of the well and take into consideration only the dependence of well
depth $\bar{u}_(\theta)$ on the orientation angle $\theta$:
\begin{equation} \label{narw32}
{\bar u}_b(\theta) \approx u_{\bf f}(r_b,\theta) \approx
u_b+fa\cos\theta
\end{equation}
with $f = |{\bf f}| > 0$, and the force effect on free diffusion in
the state outside the well. In eq. (\ref{narw31}) we took into
account the smallness of the width of the well, $a - d \ll d$, which
leads to the high accuracy of the relation $fr_b \approx fa$.

It is important to note that the small value of the well width and,
therefore, fast equilibration of the well population in radial
direction, ensures the validity of the description of the kinetics
in terms of the angular coordinate dependent well population
$n(\theta|t)$ introduced above. Noteworthy is also that the
negligible force affected change of the well shape results in the
absence of the dependence of the detailed balance relation and the
equilibrium constant $K_e (\theta)$ on the angle $\theta$. In such a
case $K_e (\theta)$ is given by the relation\cite{shu2,shu3}
\begin{equation} \label{narw32a}
K_e (\theta) = K_e^0 = a^2/Z_w,
\end{equation}
in which the effective partition function $Z_w$ is given by eq.
(\ref{nof7}), i.e. is controlled by the shape of the potential
$u(r)$ without external force, despite possible strong force effect
on the energy of the bottom predicted by eq. (\ref{narw32}). This is
because the external force leads to the identical change of both the
bottom energy ${\bar u}_b(\theta) \approx u_b+fa\cos\theta$ and the
energy of the free diffusion state at $r = a$: $u_{\bf f}(a,\theta)
\approx fa\cos\theta$.

The potential ${\bar u}_b(\theta)$ determines the kinetics of
orientational relaxation of the population in the well, which is
described by the Smoluchowski operator (\ref{narw27}) with
\begin{equation} \label{narw33}
{\bar u}(\theta)={\bar u}_b(\theta) - u_b=2\varphi\cos\theta,
\;\;\mbox{where}\; \varphi = fa/2.
\end{equation}
In this case the equilibrium state within the well is written as
\begin{equation} \label{narw34}
|\Psi_e\rangle = \frac{e^{-2\varphi\cos\theta}}{Z_{0}(\varphi)}
\;\;\mbox{with}\;\; Z_{0} (\varphi) = \frac{\sinh
(2\varphi)}{\varphi}.
\end{equation}

Substitution of formulas (\ref{narw34}) and (\ref{narw31}) into the
expression (\ref{narw30a}) yields for the function $w_e (\epsilon)$
\begin{equation} \label{narw34b}
w_e (\epsilon)/w_{e_n}^0 = Q(\epsilon) = 
\mbox{$\frac{1}{2}$}+ \varphi\coth(2\varphi) +
S(\varphi,{\epsilon}).
\end{equation}
In this formula\cite{shu2,shu3}
\begin{equation} \label{narw34bb}
w_{e_n}^0 = w_e^{0} = w_e(\varphi=0,\epsilon = 0) = Da/Z_w
\end{equation}
is the escape rate in the absence of a force and
\begin{equation} \label{narw34c}
S(\varphi,{\epsilon})= \frac{2\pi}{\varphi Z_0(\varphi)}
\sum_{l=0}^{\infty}(l+\mbox{$\frac{1}{2}$})
I_{l+\frac{1}{2}}^2(\varphi) q_l(\varphi_{\epsilon}),
\end{equation}
where
\begin{equation} \label{narw34d}
q_l (\varphi_{\epsilon}) = aq_{K_l}(\epsilon) = l+\varphi_{\epsilon}
K_{l-\frac{1}{2}}
(\varphi_{\epsilon})/K_{l+\frac{1}{2}}(\varphi_{\epsilon}).
\end{equation}
with $\varphi_{\epsilon} = \varphi\sqrt{1 + \epsilon/\epsilon_f}$
[eq. (\ref{narw18a})].

The function $w_e (\epsilon) = w_{e}^0 Q(\epsilon) $ is fairly
complicated so that, in general, the depopulation $n(t)$ can hardly
be obtained in analytical form. In the considered limit of deep
well, however, the main $\epsilon$ dependent contribution $\sim
\varphi_{\epsilon}$ comes from the term with $l = 0$ of the sum in
eq. (\ref{narw34c}). The $\epsilon$-dependence of other terms with
$l \geq 1 $, which are of higher order in $\varphi_{\epsilon}$
($\sim \varphi_{\epsilon}^2$), can be neglected by taking
$\varphi_{\epsilon} \approx \varphi_{\epsilon=0} = \varphi$. In so
doing one gets
\begin{equation} \label{narw34e}
S(\varphi,{\epsilon}) \approx \bar S(\varphi) +
\varphi^{-1}\tanh(\varphi) (\varphi_{\epsilon} - \varphi),
\end{equation}
where:
\begin{equation} \label{narw34f}
\bar S(\varphi)= S(\varphi,0)= \frac{2\pi\varphi^{-1}
}{Z_0(\varphi)}\! \sum_{l=0}^{\infty}(l+\mbox{$\frac{1}{2}$})
I_{l+\frac{1}{2}}^2(\varphi) q_l(\varphi),
\end{equation}
and
\begin{equation} \label{narw34g}
w_e (\epsilon) \approx \bar w_e +
w_{\varphi}[(1+\epsilon/\epsilon_f)^{1/2}- 1]
\end{equation}
with $\bar w_e = w_e (\epsilon = 0)= w_e^0 Q(0)$ [see eq.
(\ref{narw30a})],
\begin{equation} \label{narw34k}
\epsilon_f = (D/a^2)\varphi^2,\;\;\mbox{and}\;\; w_{\varphi} =
w_e^0\tanh\varphi.
\end{equation}
The inverse Laplace transformation of $\widetilde{\bar n}
(\epsilon)$ yields\cite{shu4}
\begin{equation}\label{narw34m}
\bar n(t) = \frac{1}{2\pi i} \int_{-i\infty+0}^{i\infty+0}
d\varepsilon \frac{\exp[\varepsilon (w t)]}{1 + \varepsilon + \bar
\gamma (\varepsilon_f + \varepsilon)^{1/2}},
\end{equation}
where $\varepsilon_f = \epsilon_f/w, \,\, w  = \bar w_0 -
w_{\varphi} = w_r + \bar w_e - w_{\varphi}$, and
\begin{equation} \label{narw34n}
\bar \gamma
=\sqrt{\frac{w_e^0}{w}}\frac{\tanh\varphi}{\varphi}\,\gamma_e
\;\;\mbox{with}\;\; \gamma_e = \sqrt{\frac{a^2 w_e^0}{D}}\,.
\end{equation}
The kinetics (\ref{narw34m}) can be expressed in terms of error
functions.\cite{shu4} In the absence of force formula
(\ref{narw34m}), naturally, reduces to that obtained earlier for
$\varphi = 0$ :\cite{shu2,shu3}
\begin{equation}
\bar n_{\varphi = 0}(t)= \frac{1}{2\pi i}
\int_{-i\infty+0}^{i\infty+0} d\varepsilon \frac{\exp[\varepsilon
(w_{0} t)]}{1 + \varepsilon + \gamma_0 \varepsilon^{1/2}},
\label{narw34p}
\end{equation}
where $w_0 = w(\varphi = 0)= w_r + w_e^0$ and  $\gamma_0 = \bar
\gamma_{\varphi = 0} = (w_e^0/w_{0})^{1/2}\gamma_e$.

Here we are not going to discuss general properties of kinetics
$\bar n(t) $ but restrict ourselves to the qualitative analysis of
its asymptotic behavior at short and long times in the limit
$\gamma_{e} \ll 1$, corresponding to the case of deep potential
well, and in the most realistic case of relatively weak force, in
which $\varepsilon_f < 1$, i.e. $D/w < 4/f^2$. In this limit at
relatively short times $t < w^{-1}\ln (1/\gamma_{\varphi}) \sim \bar
w_0^{-1}\ln (1/\gamma_0)$ the kinetics is
exponential:\cite{shu2,shu3} $\bar n (t) \approx e^{-\bar w_0 t}$
($\bar w_0 = w_r + \bar w_e$). In the opposite limit $t \gg
w^{-1}\ln (1/\gamma_{\varphi})$ the dependence $\bar n (t)$ is
non-exponential :\cite{shu2,shu3} $\bar n (t) \sim
t^{-3/2}e^{-\epsilon_f t}$. As expected, in the absence of force
($\epsilon_f = 0$) the kinetics at long times becomes of inverse
type power, which is determined by long trajectories of particles
diffusing freely outside the well.\cite{shu2,shu3} Note that for
weak force $\varphi \ll 1$ the parameter $\epsilon_f$ is small:
$\epsilon_f \sim \varphi^2$, and can be neglected. This means that
in the weak force limit with high accuracy the kinetics $\bar n(t)$
coincides with that for $\varphi = 0$ in which $w = w_0$.
Detailed analysis of specific features of the kinetics predicted by
formulas (\ref{narw34m}) and (\ref{narw34n}) is made in ref. [24].

It is easily seen that that in the considered limit of deep well the
the non-exponential inverse power tail of the kinetics $n_0 (t)$ is
small. Moreover the effect of this tail tends to zero as the value
of the force is increased. In such a case the effect of force
$\varphi$ on the escaping kinetics is reduced to that on the mean
escape rate $\bar w_{e} (\varphi) = w_e(\epsilon=0, \varphi)$
\begin{equation} \label{narw35}
\bar w_e (\varphi)/w_{e_n} = \bar Q_n(\varphi) =
\mbox{$\frac{1}{2}$}+ \varphi\coth(2\varphi) + \bar S(\varphi).
\end{equation}

Formula (\ref{narw35}) shows that the force effect on the rate is
characterized by the only parameter $\varphi = fa/2$. The numerical
calculated universal function $Q_n(\varphi)$ which describes this
effect is displayed in Fig. 2a. In addition, some limiting specific
features of the behavior of $Q_n(\varphi)$ can be revealed with
simple analytical expressions.

\begin{figure}
\setlength{\unitlength}{1cm}
\includegraphics[height=10cm,width=9cm]{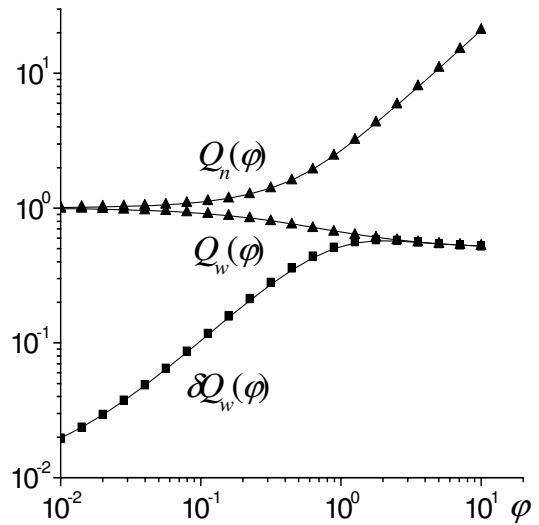}
\caption{The force ($\varphi = fa/2$) dependence of dimensionless
escape rates $Q_{\nu}(\varphi ) = w_e
(\varphi)/w^{0}_{e_{\nu}}(\varphi)$ ($\nu = n, w$) [for narrow ($n$)
and wide ($w$) wells], calculated with exact eqs. (\ref{narw35}) and
(\ref{narw45}) (full lines) and interpolation formulas
(\ref{narw41a}) and (\ref{narw50}) (triangles). Shown also is the
dependence of the dimensionless difference of rates $\delta
Q_{w}(\varphi )$ [see eqs. (\ref{dis1}) and (\ref{dis2})] calculated
with the exact formula (full line) and the interpolation expression
(\ref{narw50}) (squares). }
\end{figure}

{\it a. Weak force limit.}\, In the limit of weak external force,
when $\varphi =fa/2 \ll 1$, in two lowest orders in $\varphi$ the
mean escape rate $w_e$ can be estimated using only the first term
(with $l = 0$) in the sum $\bar S(\varphi)$ in the expression
(\ref{narw35}) for $Q_n(\varphi) = \bar w_e(\varphi)/w_{e_n}^0 $:
\begin{equation} \label{narw40}
\bar Q_n (\varphi) = \bar Q_n^w(\varphi) \approx 1 + \varphi = 1 +
fa/2.
\end{equation}

{\it b. Strong force limit.}\, In the opposite case $\varphi =fa/2
\gg 1$, denoted as the strong external force limit, the analysis of
the dependence $w_e (\varphi)$ with the use of eq. (\ref{narw35}) is
more complicated. However, the corresponding limiting dependence can
easily be derived taking into account that in the limit $\varphi \gg
1$ the escaping process is, actually, one-dimensional, for which the
$Q(\varphi)$-factor is given by\cite{shu4}
\begin{equation} \label{narw41}
\bar Q_n (\varphi) = \bar Q_n^s(\varphi)  \approx 2\varphi = fa.
\end{equation}
In deriving eq. (\ref{narw41}) we used the expression for
one-dimensional escape rate\cite{shu4} $w_1 = Df/Z_w^{(1)}$, in
which $Z_w^{(1)} = \int_{d<r<a}dr \,  e^{-u(r)} \approx Z_w/a^2 $ is
the one-dimensional partition function.

Note that this relation is also confirmed by the semiquantitative
estimation with the use of eq. (\ref{narw34f}) by truncating the sum
at $l = l^{*} \sim \sqrt{\varphi} \gg 1$ and then summing up the
terms which mainly contribute to $\bar S(\varphi)$.

{\it c. Interpolation formula.} The limiting results obtained above
for $\varphi  \ll 1$ and $\varphi \gg 1$ can be combined into a
simple algebraic interpolation formula
\begin{equation} \label{narw41a}
Q_n(\varphi) \approx Q_n^{(i)}(\varphi) = 1 + \varphi
(2-e^{-\varphi})\end{equation} which reproduces function $Q_n
(\varphi)$, numerically evaluated using eqs. (\ref{narw34f}) and
(\ref{narw35}), with accuracy $\sim 3\%$ (see Fig. 2a).

\subsubsection{Wide-well shape}

Another form of the well shape, in which analysis of the escape rate
$w_e$ can be made analytically, corresponds to the small distance of
closest approach, or large $a$, for which $d \sim r_b \ll a$. In
this case in a fairly wide region of relatively strong force $f <
1/d,1/r_b $ the escape kinetics is fairly accurately described by
the TSM (\ref{prf7b}).

It is important to note that the inequality $fr_b < 1$ ensures quite
high accuracy of the approximation neglecting the effect of force on
the well shape in the region near the bottom. In this approximation,
the quasiequilibrium population distribution within the well is
isotropic:
\begin{equation} \label{narw42}
|\Psi_e\rangle = |0\rangle.
\end{equation}
This, in turn, means that the partition function $Z_w$ is
independent of the angle $\theta$ and is given by eq. (\ref{nof7}).

The effect of force, however, manifests itself in the anisotropy of
the activation energy of escaping $u_a (\theta)$:
\begin{equation} \label{narw43}
u_a (\theta) \approx u_{\bf f} (\theta, a) \approx u_b + 2\varphi
\cos \theta , \;\;\;(\varphi  = fa/2),
\end{equation}
which, in turn, leads to the anisotropy of the detailed balance
relation, i.e. the anisotropy of the equilibrium constant
\begin{equation} \label{narw44}
K_e (\theta) = K_e^0 e^{-2\varphi\cos\theta},
\end{equation}
where $K_e^0$ is the isotropic equilibrium constant in the absence
of external force given by eq. (\ref{narw43}).

Formula (\ref{narw44}) calls for some additional comments especially
concerning its applicability. The fact is that the value of $K_e
(\theta)$ at each particular $\theta$ is determined assuming local
quasiequilibrium of the population outside and inside the well in
the region close to $r = a$ at this $\theta$. In general, it is
difficult to justify the existence of the quasiequilibrium in the
considered limit, unlike the limit of narrow well discussed above.
This is because for $\varphi = fa/2 \lesssim 1$ the time of passing
over the escaping barrier width  $\delta_b \sim {\rm min}\{a,
f^{-1}\} $ (the width of the region of transition from the inner
part of the well to the outer one), $\tau_{b} \sim \delta_b^2/D$ is
comparable with the time of reorientation $\tau_c \sim a^2/D$. It is
worth noting, however, that the accuracy of quasiequilibrium
assumption becomes better with increasing $f$ since the for $\varphi
= fa/2 \gg 1$ the width $\delta_b \ll a$ and, correspondingly,
$\tau_{b} \ll \tau_c$.

The above-mentioned arguments lead us to the conclusion that in the
considered limit of small radius of the well bottom the TSM with
$\theta$-dependent equilibrium constant $K_e (\theta)$ gives quite
reasonable interpolation formula for the kinetics of the escaping
process and, in particular, for the escape rate $w_e$, which
correctly describes both the limit of weak and strong external
force. Further analysis (see below) will confirm this statement.

Formula for the escaping kinetics can straightforwardly be derived
with the use of general formulas (\ref{narw30}), (\ref{narw30a}),
and some results obtained above in the limit of narrow potential
well. The fact is that, in the mathematical form, the average of any
operator multiplied by angular dependent equilibrium constant
(\ref{narw44}) [of type of eq. (\ref{narw30a})] over the isotropic
equilibrium state (\ref{narw42}) is similar to the average over the
equilibrium distribution (\ref{narw34}), except for the partition
function $Z_0 (\varphi)$ [eq. (\ref{narw34})], which should be
replaced by $Z_0 (\varphi\to 0) = 2$ corresponding to the isotropic
distribution. These simple algebraic manipulations result in the
following expression for $w_e (\varphi)$
\begin{equation} \label{narw45}
\bar w_e (\varphi)/w^0_{e_{w}}(\varphi) = \bar Q_{w}(\varphi) =
\mbox{$\frac{1}{2}$}Z_0(\varphi)e^{-2\varphi} \bar Q_{n}(\varphi)
\end{equation}
where $Z_0 (\varphi)$ and $Q_{n}(\varphi)$ are determined in eqs.
(\ref{narw34}) and (\ref{narw35}), respectively, and
\begin{equation} \label{narw47}
w^{0}_{e_w}(\varphi) = w_e^{0} e^{2\varphi}
\end{equation}
is the escape rate in the absence of the external force but with the
activation energy $u_a^{*}$, corresponding to the orientation
$\theta = \pi$ (most favorable for escaping):
\begin{equation} \label{narw47a}
u_a^{*} \equiv  u_a (\theta = \pi)   = u_b - 2\varphi.
\end{equation}

As in the case of narrow potential well the dependence of $w_e$ on
the force $f$ is expressed in terms of that on the only parameter
$\varphi$. The characteristic function $Q_{w}(\varphi)$, which
determines the pre-exponential factor in the activation type
dependence of $w_e (\varphi)$, is displayed in Fig. 2b.  The
numerical results show that $Q_w (\varphi)$ monotonically decreases
(with increasing $\varphi$) from $Q_w  = 1$ at $\varphi = 0$ to $Q_w
= 1/2$ at $\varphi \to \infty$. This behavior is markedly different
from that of $Q_n (\varphi)$ although the specific features of $Q_w
(\varphi) $-dependence are essentially determined by those of $Q_n
(\varphi)$. Some of features of the function $Q_w (\varphi) $, for
example saturation at $\varphi \to \infty$, looking unexpected at
first sight, can be understood by simple analysis (see below).

{\it a. Weak force limit.}\, In the weak force limit the behavior of
$Q_w (\varphi)$ at $\varphi =fa/2 \ll 1$ differs form that obtained
above for $Q_n (\varphi)$ (i.e. for narrow potential well):
$Q_w(\varphi)$ decreases with increasing $\varphi$, so that at
$\varphi \ll 1$
\begin{equation} \label{narw48}
\bar Q_w^w(\varphi) = \bar Q_w (\varphi\ll 1) \approx 1 -  \varphi.
\end{equation}
Such a behavior of $Q_w(\varphi)$  results from using the
$\varphi$-dependent normalizing rate $w^0_{e_w} \sim e^{2\varphi}$
(instead of $w_{e_n}^{0} = w_e^{0}$) in the definition of
$Q_w(\varphi)$.

{\it b. Strong force limit.}\,  In the opposite limit $\varphi =fa/2
\gg 1$ the force strongly affects the average escape rate $w_e$,
first of all, because it significantly changes the activation energy
of the rate $w_e$. As for $Q_w (\varphi)$, which characterizes the
pre-exponential factor of the corresponding Arrenius-type expression
for $w_e$, at $\varphi \gg 1$ it monotonically decreases approaching
the asymptotic value $1/2$.

The obtained $Q_w (\varphi) $-independence at $\varphi \to \infty$
can easily be understood by taking into account that, according to
formula (\ref{narw44}), in the case of wide well for strong external
forces the flux of escaping particles is highly localized in a small
region of orientations $\delta \theta = \pi - \theta \lesssim
1/\sqrt{\varphi}\ll 1$. The escape rate is determined by the total
flux $J_e$ through this region of size $s_e \sim (\delta\theta)^2
\sim \varphi^{-1}$. In the strong force limit $\varphi \gg 1$ the
flux $J_e \sim \varphi$, as it follows from eq. (\ref{narw41}), so
that $Q_w (\varphi) \sim s_e J_e \sim {\rm const}$. The exact
estimation can be obtained just by substitution of the corresponding
limiting expression (\ref{narw41}) into eq. (\ref{narw45}):
\begin{equation} \label{narw49}
\bar Q_w^s (\varphi)=\mbox{$\frac{1}{2}$}Z_0(\varphi)e^{-2\varphi}
\bar Q_{n}^s(\varphi)|_{\varphi\gg 1} = \mbox{$\frac{1}{2}$}
\end{equation}

Similarly to the narrow well limit, in the case of wide well for
large $\varphi$ the escape rate is determined by the
quasi-one-dimensional flux of escaping particles. The mechanism of
formation of the one-dimensional flux is, however, somewhat
different in both cases: for narrow wells the transition to the one
dimensional regime results from high localization of the well
population in the small region at $\theta \sim \pi$, while for wide
wells this transition is caused by strong localization of favorable
transition rates in this region.

{\it c. Interpolation formula.} A simple interpolation expression
for $Q_w (\varphi)$ can be derived, for example, with the use of
similar formula for $Q_n (\varphi) $ presented in eq.
(\ref{narw41a}):
\begin{equation} \label{narw50}
\bar Q_w (\varphi)\approx \bar Q_w^{i}
(\varphi)=\mbox{$\frac{1}{2}$}Z_0(\varphi)e^{-2\varphi} \bar
Q_{n}^i(\varphi).
\end{equation}
Quite satisfactory accuracy ($\sim 3\%$) of this formula is
demonstrated in Fig. 2b.

\section{Escaping kinetics in 2D case}

In this section we briefly discuss the kinetics of force affected
two dimensional escaping of Brownian particles from the well.

The formal analysis of the 2D problem is very similar to that in the
3D cases made above. The differences reduce to a few changes of
analytical forms of dependences on the parameters of the model.

First, the 2D TSM equations are similar to 3D eqs. (\ref{prf7}) but
with 2D variants of parameters and operators: $S_a = 2\pi a$, $\hat
L_{f} = D\nabla_{\bf r}(\nabla_{\bf r} + {\bf f})$ is the operator
describing 2D diffusive motion outside the well, and $\hat L_c = D_c
\nabla_{\theta}(\nabla_{\theta} + \nabla_{\theta}u_a(\theta))$ is
the operator of orientational (circular) diffusion in the circular
cage in 2D space. The analytical form of the equilibrium constant
$K_e (\theta)$ is also somewhat different from those discussed above
in 3D case (see below).\cite{shu2,shu3}

Similar to 3D case the orientation dependent population of the well
$n(\theta|t)$ and PDF of particles outside the well $c(r,\theta|t)$
are conveniently represented as vectors by expanding them in basis
of vectors
\begin{equation}
|l\rangle = N_l\cos(l\theta), \;\,\langle l| =\!
\int_{-\pi}^{\pi}\!\!d\theta \cos(l\theta)\! \dots,\;\,(l\geq 0),
\label{twod1}
\end{equation}
where $N_{l\geq 1} = 1/\pi$ and $N_0 = 1/(2\pi)$.

The solution of (2D) TSM equations for vectors $|{\bf n} (t)\rangle$
and $|{\bf c} (r|t)\rangle$ by the method applied in the 3D case
yields for $\tilde n_0 (\epsilon) = \int_{-\pi}^{\pi}\!d\theta\,
\widetilde{n}(\theta,\epsilon) \equiv \langle 0|\widetilde{\bf
n}(\epsilon)\rangle$ the expression (\ref{narw8a}). In the fast
orientational relaxation limit this expression, naturally, reduces
to formula (\ref{narw30}). In both these equations the operator
$\hat W_e (\epsilon)$ is given by eq. (\ref{narw4}) in which
\begin{equation} \label{twod3}
\hat G^{-1}(a,a|\epsilon) = D[e^{-\varphi\cos\theta}\hat q_K
(\epsilon) e^{\varphi\cos\theta} - \mbox{$\frac{1}{2}$}f
\cos\theta],
\end{equation}
where $\hat q_K(\epsilon) = \sum_{l=0}^{\infty}|l\rangle
q_{K_l}(\epsilon) \langle l| $ with
\begin{equation} \label{twod4}
q_{K_l}(\epsilon) = a^{-1}\big[l + \varphi_{\epsilon} K_{l-1}
(\varphi_{\epsilon})/K_{l} (\varphi_{\epsilon})\big].
\end{equation}

With the use of thus obtained general expressions and the
relation\cite{Abr}
\begin{equation} \label{twod5}
e^{\pm\varphi\cos\theta} =  I_0 (\varphi) +
2\sum\nolimits_{l=1}^{\infty} \!I_{l}(\varphi)\cos(l \theta)
\end{equation}
one can derive relatively simple formulas for the depopulation
kinetics $n_0 (t)$ within two models of the well shape discussed in
Sec. IIID: the models of narrow and wide well.

It is worth noting that the kinetics of 2D diffusion-assisted
reactions are essentially different from that of 3D processes
because 2D diffusion is recurrent while 3D one is transient. In the
particular case of diffusion-assisted escaping from the well the
recurrent nature of 2D diffusion manifests itself in strongly
non-exponential escaping kinetics even in the case of deep well.
Moreover in the absence of external force $\bar w_e = w_e (\epsilon
= 0) = \langle \Psi_e |\hat W_e (\epsilon = 0)|\Psi_e\rangle = 0$,
i.e. the probability of escaping from the well is zero and escaping
is always reversible (which is quite natural for recurrent process).

This, in turn, means that unlike 3D escaping processes, in the
absence of force, the efficiency of 2D ones cannot be characterized
by the only parameter. The external force, however, results in the
irreversible escaping flux which shows itself in $\bar w_e (\varphi
\neq 0) \neq 0$ and nearly exponential escaping kinetics approaching
the exponential with the increase of force. Below we will restrict
ourselves to the evaluation the mean escape rate $\bar w_e
(\varphi)$ describing the force effect on the escape process, which,
as we have already mentioned above, is getting closer to exponential
with increasing the force ($\varphi$).

\subsection{Narrow-well shape}

In the 2D case of narrow wells $a -d \ll d$, which is defined
similarly to the 3D case (see Sec. III.E.1.), in the large region of
forces $f$ the force effect reduces to the change of the
quasiequilibrium distribution within the well. The corresponding 2D
quasiequilibrium  state $|\Psi_e\rangle$ can be written as
\begin{equation} \label{twod6a}
|\Psi_e\rangle = I_{0}^{-1}(2\varphi)e^{-2\varphi\cos\theta}.
\end{equation}
With the use of this expression and formulas, derived above, we get
\begin{equation} \label{twod6}
\bar w_e (\varphi)/w_{e_n}^* = \bar Q_n(\varphi) = \varphi
I_1(2\varphi)/ I_0(2\varphi) + \bar S_p(\varphi),
\end{equation}
where
\begin{equation} \label{twod7}
w_{e_n}^{*} = D/Z_w^{(2)}, \;\;\mbox{with}\;\;Z_w^{(2)}=
\int_{d<r<a}\!\!dr \, r e^{-u(r)},
\end{equation}
is the auxiliary parameter of dimensionality of rate and
\begin{equation} \label{twod8}
\bar S(\varphi) = \varphi I_0^{-1}(2\varphi)
\sum\nolimits_{l=0}^{\infty}\!\zeta_l I_{l}^2(\varphi) K_{l-1}
(\varphi)K_{l}^{-1} (\varphi),
\end{equation}
with $\zeta_0 = 1$ and $\zeta_{l\geq 1} = 2$.

The numerically calculated dependence $\bar w_e (\varphi)$
(\ref{twod6}) is displayed in Fig. 3. As expected $\bar w_e (\varphi
\to 0) \to 0$. In the opposite limit $\varphi \gg 1$ we get the
dependence
\begin{equation} \label{twod8a}
\bar w_e (\varphi \gg 1)/w_{e_n}^* \approx 2\varphi = fa
\end{equation}
similar to that obtained in the 3D case and corresponding to the 1D
escaping mechanism (see Fig. 3). To clarify this statement note that
for narrow wells $Z_w^{(2)} \approx aZ_w^{(1)}$, where $Z_w^{(1)} =
\int_{d<r<a}\!dr \, e^{-u(r)} $ is the 1D partition function, and
therefore $\bar w_e (\varphi\!\gg\! 1) \approx 2\varphi w_{e_n}^* =
fa(D/Z_w^{(2)}) \approx Df/Z_w^{(1)}$.

\subsection{Wide-well shape}

The opposite limit of (2D) wide wells ($a -d \gg d$), defined by
analogy to the 3D case (Sec. III.E.2.), can be treated with the use
of formulas obtained in the narrow-well case. To do that one should
take into account, similarly to the 3D case, that in the wide-well
limit $|\Psi_e\rangle = |0 \rangle$ and $K_(\theta) \sim
e^{-2\varphi \cos \theta}$. This means (as it is mentioned in Sec.
III.E.2) that both limits the average $w_e (\epsilon) = \langle
\Psi_e\rangle| \hat W_e (\epsilon) |\Psi_e\rangle$ is similar from
mathematical point of view and differ only of the normalization
factor $I_0^{-1} (2\varphi)$
\begin{equation} \label{twod10}
\bar w_e (\varphi)/w_{e_w}^* = \bar Q_w(\varphi) =\bar
Q_n(\varphi)I_0 (2\varphi)e^{-2\varphi},
\end{equation}
where
\begin{equation} \label{twod11}
w_{e_w}^{*} = w_{e_n}^{*}e^{2\varphi}.
\end{equation}

The dependence $\bar w_e (\varphi)$ (\ref{twod10}) is shown in Fig.
3. In agreement with the above statement in the limit of weak force
$\bar w_e (\varphi \to 0) \to 0$. In the opposite limit of strong
force $\varphi \gg 1$, however, one gets $\bar w_e (\varphi) \sim
\sqrt{\varphi}$ as it follows from the analysis of the numerically
calculated function $\bar w_e (\varphi)$ presented in Fig. 3. The
dependence $\bar w_e (\varphi \gg 1) \sim \sqrt{\varphi}$ can easily
be understood by estimations similar to those, clarifying the $\bar
w_e (\varphi)$-independence of $\varphi$ at $\varphi \gg 1$ (see
Sec. III.E.2). The fact is that for $\varphi \gg 1$ the density
escaping flux $J_e \sim \varphi$ is localized is localized in the
small sector of size $\delta\theta \sim 1/\sqrt{\varphi}$ around the
most favorable direction $\theta = \pi$. Therefore the total
escaping flux $Q_w \sim J_e \delta\theta \sim \sqrt{\varphi}$. The
exact asymptotic dependence $\bar w_e (\varphi \gg 1)$ can be
obtained with eq. (\ref{twod10}) and formula (\ref{twod8a}):
\begin{equation} \label{twod12}
\bar w_e (\varphi \gg 1)/w_{e_w}^{*} \approx \sqrt{\varphi/\pi}.
\end{equation}
High accuracy of this expression at $\varphi \gtrsim 1$ is
demonstrated in Fig. 3.

\section{Discussion and applications}

\subsection{General remarks}

This work concerns detailed theoretical study of the effect of the
external force $f = F/(k_B T)$ on the kinetics of diffusion-assisted
depopulation of a deep isotropic potential well in 2D and 3D
processes. Fairly simple matrix expressions for the depopulation
kinetics are obtained and thoroughly analyzed.

In our work we have concentrated on the analysis in the most
physically reasonable limit of fast orientational relaxation of the
population in the well. In this limit the analytical expression for
the depopulation kinetics is derived which predicts the kinetics to
be close to the exponential in the wide region of parameters of the
model both for 2D and 3D processes. The total depopulation rate in
this case is shown to be a sum of reaction and escape rates. In our
work we have mainly studied the specific features of the escape rate
$w_e$ whose value appears to significantly depend on shape of the
well. Simple analytical expressions for $w_e (f)$ are obtained for
two limiting types of wells: narrow wells of type of well layer at a
distance of closest approach $d$ (for which $a-d \ll d$) and wide
wells with large effective Onsager radius $a \gg d$.

In the case of narrow well the effect of the force on the escape
rate is fairly strong but shows itself only in the preexponential
factor of the Arrenius-type dependence of the rate, i.e. no strong
effect on the activation energy is predicted. On the contrary, in
the case of wide well (or small distance $d$) the force affects not
only preexponential factor but the activation energy as well.

It is worth noting that the effect of an external force on the
diffusion-assisted processes in the presence of interaction between
particles are studied in a number of works (see, for example, refs.
[1] and [7]). Especially comprehensively the force effect (electric
field effect) is analyzed in the case of ion pair recombination
reaction, i.e. in the case of the Coulomb interaction between
particles.

Unfortunately it is practically impossible to compare the results of
our analysis with majority of those obtained earlier, since these
works mainly concerned with processes in potentials without well at
short distances, the reactivity is usually assumed to be high. In
particular, in the case of ion pair recombination processes the
recombination kinetics is considered to be determined by diffusive
motion in the pure Coulomb potential.\cite{Ri,Noo,Noo1} It is,
nevertheless, interesting to note that in the small field limit $fa
\ll 1$ the force effect on the probability $P_e (f)$ of escape from
the Coulomb potential, found in ref. [5], is independent of the
initial distance between ions and is represented in the form $P_e
(f) \approx P_e (f=0)(1 + fa/2)$, which is in apparent agreement
with the field dependence of the escape rate obtained in our work
[see eqs. (\ref{narw40}) and (\ref{narw48})].

Noteworthy is also that in the strong force limit the escaping
process becomes nearly one-dimensional in both cases of well shape
considered in both for 2D and 3D processes. In this limit the escape
rate is determined by the 1D flux in the small region of favorite
orientations corresponding to $\theta \sim\pi $. This fact allows
one to easily improve the considered TSM, in which the effect of the
force on the location of top of the barrier (assumed to be at $r =
a$) is neglected. Moreover, in the strong force limit one can also
take into account the smoothness of the shape of the realistic
barrier near the top, which in the TSM is actually assumed to be of
cusp shape.

Concluding this short discussion of results we would like to note
that in this work we restricted ourselves to the analysis of the
most realistic limit of fast orientational relaxation within the
well. In reality, however, with the use of general formula
(\ref{narw8a}) one can also describe the manifestation of finiteness
of the orientation relaxation time. The case, in which the effect of
finiteness is largest, of course, corresponds to  $\hat L_c = 0$,
i.e. the absence of orientational relaxation. In this case the
angular dependence of the equilibrium rate [$K_e (\theta)$],
evidently, results in the highly non-exponential well depopulation
kinetics $n_0 (t)$, which can be approximated by the sum of
exponentially decreasing (monomolecular) contributions with
$\theta$-dependent rates, coming from different orientations. With
the use of obtained formulas there will be no difficulties to
analyze this case as well, when needed.

In our further discussion possible applications of formulas obtained
we will consider some types of processes recently actively analyzed
experimentally.

\subsection{Condensed phase reaction kinetics.}

The obtained formulas are quite suitable for the analysis of
diffusion-assisted condensed phase geminate reactions. The effect of
external force on escaping and reaction yields is of special
interest in the particular case of reaction ion-pair recombination
reactions in polar solids and liquids, in which the external force
can be realized by applying electric field.  There are a number of
experimental time resolved spectroscopic studies of reactions with
the participation of ions.\cite{Chen,Bart1,Bart2,Shkrob,Crow}

Traditionally the results of such investigations are analyzed with
the use of model calculations in which interaction is assumed to be
pure Coulomb [.]. It is worth noting, however, that in polar media
the medium affected interaction, which can be described by the MFP,
strongly deviates from the from the Coulomb one [.]: unlike the
Coulomb potential the MFP oscillates at short distances of order of
molecular size [.]. At distances close to that of the first
coordinate shell the MFP has a most deep well, whose depth can be
much larger than $k_B T$ for dielectric constants $\epsilon \gtrsim
20$. In this case the proposed approximation of the realistic
interaction with the use of well-type potential is much more
accurate than pure Coulomb potential approach.

Some of earlier results of the proposed TSM, concerning the kinetics
of processes in the absence of external force ($f = 0$), have
already been successfully applied to the analysis of kinetics of
some liquid-phase reactions.\cite{Ri,Street1,Que,Woj} Concerning the
results obtained above for the case $f \neq 0$, note that the most
convenient for experimental analysis is not the force dependent
inverse mean lifetime ${w}_0 (\varphi)$ [see eq. (\ref{narw30a})],
but the difference ${w}_0 (\varphi) - {w}_0 (0)= w_e (\varphi) -
w_e^0$, which is independent of the rate $w_r$ of reaction within
the well (assumed to be independent of $\varphi$). The corresponding
dimensionless parameters
\begin{equation} \label{dis1}
\delta Q_{\nu} (\varphi)= [{w}_0 (\varphi) - {w}_0
(0)]/w_{e_{\nu}}^{0} (\varphi), \;\;(\nu = n, w),
\end{equation}
are directly related to $Q_{\nu} (\varphi)$:
\begin{equation} \label{dis2}
\delta Q_{n} (\varphi) =  Q_{n} (\varphi) - 1,\;\; \delta Q_{w}
(\varphi) =  Q_{w} (\varphi) - e^{-2\varphi}.
\end{equation}
The behavior of $\delta Q_{n} (\varphi)$ is, clearly, similar to
that of $Q_{n} (\varphi)$ except for evident displacement along
ordinate axis. As for $\delta Q_{w} (\varphi)$-dependence, shown in
Fig. 3, its form is essentially different from that of $Q_{w}
(\varphi)$: at $\varphi \to 0$ the function $\delta Q_{w} (\varphi)
\approx \varphi$ is similar to $\delta Q_{n} (\varphi)$, while
$\delta Q_{w} (\varphi \to \infty) = 1/2$. Moreover $\delta Q_{w}
(\varphi)$ has a maximum (though not very pronounced) at $\varphi =
\varphi_m \approx 2.0$.

It is of great interest to compare these theoretical predictions
with experimental results of type of those given in refs. [9-13] but
in the presence of electric field.

\subsection{Transient photocurrents.}

Recently considerable interest has been attracted to experimental
investigations of the recombination kinetics of ion pairs in
non-polar and moderately polar liquids with time resolved
measurements of transient photocurrent
\cite{Zhou1,Zhou2,Zhou3,Lukin}. The photocurrent is found to be
fairly sensitive to the spatial evolution of photoinduced ion pairs
and, in particular, to the recombination kinetics.

The proposed theory can be very fruitful for the interpretation and
description of experiments on transient photocurrents. The fact is
that the TSM considered above enables one to quite accurately
describe the spatial evolution of ion pairs without solving the
Smoluchowski equations which are rather complicated even for
restrictive models of the interparticle interaction. In this short
discussion we will outline and illustrate the possibilities of the
proposed method.

Recall that the TSM is valid in the limit of the time $\tau_c$ of
the PDF relaxation in the well much shorter than the average
lifetime $\bar \tau_0 = \bar w_0^{-1}$ in the well [according to eq.
(\ref{narw30}) $ w_0^{-1}\sim (w_e^0)^{-1}, w_r^{-1}$]. Noteworthy
is also that the TSM describes the kinetics at relatively long times
$t > \tau_c$. In reality, however, the method can straightforwardly
be extended to also treat the PDF relaxation at short times $t
\lesssim \tau_c$ (i.e. relaxation within the well) just because of
assumed essential difference between thermalization time $\tau_c$
and $\tau_0$. This difference leads, in fact, to the independence of
the processes of intrawell relaxation and well depopulation, which
can be described separately.

The initial stage of relaxation of the ion-pair PDF within the well
results in the transient photocurrent $J_e (t)$ at times $t \lesssim
\tau_c$. The relaxation kinetics depends on the well shape. Below,
for simplicity we will consider the case of narrow well and assume
that the initial PDF in the well is isotropic. For narrow wells the
intrawell PDF relaxation can, in turn, be separated into two stages:
radial and orientational relaxation. The first stage is the radial
relaxation whose characteristic time is $\tau_w = \Delta_w^2/D$,
where $\Delta_w = a-d < d$, is shorter than the time $\tau_c =
D_c^{-1} \approx r_b^2/D$ of the second stage of orientational
relaxation to the anisotropic quasiequilibrium ion-pair PDF in the
well of the potential $u_{\bf f} ({\bf r})$ (\ref{form2}).

Together with the final stage, described by the TSM, the kinetics of
the transient current relaxation can be represented as follows:

1. At $t \sim \tau_w $ (the stage of radial relaxation in the well)
the transient current can approximately be described by the simple
expression
\begin{equation} \label{dis3}
J_e (t) \approx \bar J_0 + (J_0 - \bar J_0) e^{-t/\tau_w},
\end{equation}
where $J_0 = eDf $ is the current produced by freely diffusing ion
pair [initial current for the isotropic initial PDF (\ref{form0})]
and $\bar J_0 \mbox{$\frac{2}{3}$}J_0$ is the final current after
radial relaxation [see eq. (\ref{dis4}]

2. At $t \sim \tau_c$ (the stage of orientational relaxation in the
well) the current $J_e (t) $ can be evaluated using the relation
$\bar J_e = \dot {\bar p}(t)$, where ${\bar p}(t) = e\langle r_f(t)
\rangle $ is the the average electric dipole moment of ion pairs
within the well, in which where $r_f$ is the projection of radius
vector of the ion pair on the direction of the field: $r_f = ({\bf
r}\cdot {\bf f}/)f$. For the case of narrow well the derivative of
the average dipole moment $\dot {\bar p}(t)$ is easily evaluated in
the limit of weak external force $f$ in the lowest order in $f$ (in
the linear response approach):
\begin{equation} \label{dis4}
J_e (t) \approx \bar J_0e^{-2D_c t}\;\; \mbox{with}\;\; \bar J_0 =
\mbox{$\frac{2}{3}$}J_0,
\end{equation}
where $\,J_0 $ is defined in eq. (\ref{dis3}), and $D_c \approx
D/r_b^2$ ($r_b$ is the radius of the well bottom).

3. At $t > \tau_c$ (the stage escaping from the well) the TSM
predicts the following expression for the current:
\begin{equation} \label{dis5}
J_e (t) \approx J_0 n_e (t) \;\;\mbox{with} \;\; n_e(t) = n_s(t) -
n(t).
\end{equation}
Here
\begin{equation} \label{dis7}
 n_s (t) =1 - w_r \int_0^{t}\! d \tau \, n(\tau)
\end{equation}
is the ion-pair survival probability.

Taking into account different time scales of these three kinetic
stages one can describe the current relaxation kinetics by the
combined expression
\begin{equation} \label{dis8}
j_e(t) = J_e (t)/J_0 \approx \mbox{$\frac{1}{3}$} (2 +
e^{-t/\tau_w})e^{-2D_c t} + n_e (t).
\end{equation}

This simple expression enables one to quantitatively describe the
specific features of the current relaxation kinetics in a wide
region of times. In deriving eq. (\ref{dis8}) we have assumed the
well to be narrow. As applied to the ion pair recombination and
transient current relaxation this assumption is quite appropriate in
the limit of relatively polar solvent with dielectric constant
$\epsilon_s \gtrsim 15$, in which the ion-pair MFP $u(r)$ is known
to be of the shape of deep narrow well at short distances $r \sim
d$.\cite{Pat,Nin}

Formula (\ref{dis8}), however, turns out to be of quite reasonable
accuracy even in the case of diffusion-assisted processes in the
pure Coulomb potential $u (r) = a/r $ with nearly reflective (low
reactivity) boundary condition at $r = d$, whose well (at $r \gtrsim
d$), at first sight, can hardly be treated as narrow for $a \gg d$.
The fairly good accuracy of eq. (\ref{dis8}) in this case results
from the cusp-like shape of the well at $r\sim d$, where the major
of the well population is localized (see below).

To demonstrate the accuracy of eq. (\ref{dis8}) we will compare its
predictions with recent calculations of the current relaxation
kinetics based on numerical solution of the Smoluchowski
equation.\cite{Lukin} These calculations have been made for $u (r) =
a/r $, with $a = 63.3$ A, and for small force $\varphi = 0.05$,
which only very weakly affects the escaping rate $\bar w_e$. The
other parameters of the model are:\cite{Lukin} the coefficient $D =
3.1 \cdot 10^{-5} \,{\rm cm^2/s}$ of relative diffusion of ions, the
contact distance $d = 9$ A, and the reactivity $\sigma_r$, which
determines the reactive flux at a contact distance $r = d$: $ D
[\nabla_r \rho + (\nabla_{r} u +f\cos\theta)\rho]|_{r=d}
=\rho(\sigma_r/4\pi d^2)|_{r=d}$, and whose value $\sigma_r =
0.01\cdot(4\pi D a)$ corresponds to the weak reactivity limit.

Recall that the discussed TSM is applicable in the limit of small
$\gamma_e =(a^2 w_e^0/D)^{1/2} \ll 1$  [see eq. (\ref{narw34n})].
However, for the parameters of the model chosen above $\gamma_e
\approx 0.9$, i.e. the TSM is not expected to reproduce the exact
numerical results very accurately. Nevertheless, even in this case
the accuracy of this model appears to be quite good (see below).

Noteworthy is also that in the considered model assuming not very
deep well of cusp-like shape the quasiequilibrium PDF in the well is
somewhat displaced to $r > d$. In this case, with reasonable
accuracy one can put $r_b = \bar r = \int_d^{\infty}dr \, r
e^{-u(r)}/Z_w \approx 19 \,{\rm A}$ so that $D_c \approx
8\cdot10^{8} \, {\rm s}^{-1}$ [eq. (\ref{dis4})]. For this value of
$r_b$ one can also approximately estimate $\tau_w^{-1}$:
$\tau_w^{-1} \approx D/(r_b - d)^2 \approx 3.1\cdot 10^{9} \,{\rm
s}^{-1}$.

For the chosen parameters of the model the survival probability
$n_s^{\infty} = n_s (t\to\infty) \approx  0.085$.\cite{Lukin} Taking
into account that the TSM exactly predicts the asymptotic (at $t \to
\infty$) behavior of the escaping kinetics we can write the relation
$n_s^{\infty} = \bar w_e/ \bar w_0 = \bar w_e/(w_r + \bar w_e) =
0.085$.

\begin{figure}
\setlength{\unitlength}{1cm}
\includegraphics[height=7.1cm,width=9.3cm]{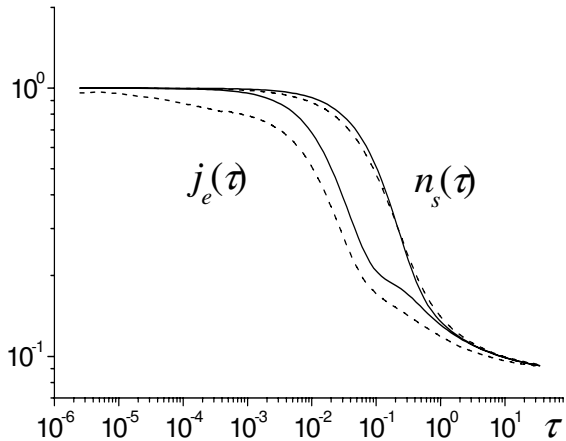}
\caption{The comparison of time ($\tau = tD/a^2$) dependences of the
ion-pair survival probability $n_s (\tau)$ and the normalized
transient current $j_e (\tau) = J_e (\tau)/J_0$, calculated
analytically with eqs. (\ref{dis7}) and (\ref{dis8}) (full lines)
and numerically\cite{Lukin} (dashed lines). The parameters used are
presented in Sec. V.C. }
\end{figure}

Figure 3 displays the comparison of the time dependent survival
probability $n_s (t)$ and the transient current $J_e (t)$ with those
calculated by numerical solution of the Smoluchowski
equation.\cite{Lukin} For convenience of the comparison with
numerical results in Fig. 3 we use the dimensionless time $\tau =
Dt/a^2$. The comparison shows good accuracy of the TSM prediction
for the kinetics $n_s (t)$, which is determined by the evolution of
the system at long times $\tau >  0.1$. As for the time dependence
of the transient current $J_e (\tau)$, it is also quite accurately
reproduced at long times, when the contribution of $n_s(t)$-kinetics
[represented by the last term in eq. (\ref{dis8})] is dominated.
Some (not very strong) difference between analytical and numerical
behavior at shorter times $\tau < 0.1 $ is a result of
above-mentioned approximations applied in deriving eq. (\ref{dis8})
for $J_e(t)$ at first two short time stages of the current
relaxation.

\subsection{Optical tweezers.}

The important problem which have recently attracted much attention,
and in study of which the TSM can be very suitable is the kinetics
of trapping of colloidal particles by optical tweezers, i.e. by the
optical force induced potential wells (arising due to gradients of
laser beam intensity).\cite{Ashkin,Grier}  Detailed experimental
investigations show that the tweezers potential well $u(r)$ is
highly localized (the size is about $\mu$m ) and is expected to be
highly localized.\cite{Via} Of course, in general the well is not
spherically symmetric. Moreover, recently the axial (non-potential)
component of the trapping force is found.\cite{Grier1} However, to a
good accuracy, for description of trapping and escaping kinetics one
can neglect these non-trivial contributions to the trapping force
and use the isotropic approximation for the tweezers well.

The proposed theory based on the TSM is very useful for the analysis
of the kinetics of trapping into and escaping from tweezers
potential well. Application of the TSM significantly simplifies the
problem of description of the kinetics thus allowing for the
analysis of more complicated effects of a large number potential
wells.\cite{Grier2}

Of special interest is the effect of well motion on the
trapping/escaping kinetics discussed in a number of papers
concerning possible applications of tweezers.\cite{Evst} This effect
is known to reduce to that of the external force discussed above.
The fact is that the well motion induces the force ${\bf F}$ acting
on a particle. In the frame of reference, moving with the well, the
force is proportional to the well velocity ${\bf v}$: ${\bf F}_v =
\mu^{-1} {\bf v}$, where $\mu = D/(k_B T)$ is the mobility of the
particle, i.e ${\bf f}_v = D^{-1}{\bf v}$. Note that this relation
holds in the case of time dependent velocity ${\bf v} (t)$ as well.

Thus the problem of the analysis of the effect of well motion is
equivalent to that thoroughly discussed above with the external
force  ${\bf f} = {\bf f}_v$ (and without reactivity, i.e with $w_r
= 0$). In our consideration we have assumed that ${\bf f}$ is
independent of time. The case of time dependent ${\bf f}$ [for
example, because of time dependent velocity ${\bf v} (t)$] is, in
general, much more complicated and can hardly be analyzed
analytically. However, in a quite realistic case of relatively
slowly changing force, for which the characteristic changing time
$\tau_f > 1/\bar w_0$, one can treat the force effect adiabatically
evaluating the nearly exponential escaping kinetics with the use of
formulas derived above for static $f$, in which the escaping rate
$\bar w_e (\varphi)$ is replaced by the corresponding time dependent
expression $\bar w_e (\varphi(t))$.

\textbf{Acknowledgements.}\, The author is grateful to Dr. V. P.
Sakun for valuable discussions. The work was supported by the
Russian Foundation for Basic Research.




\begin{thebibliography}{99}
\bibitem{Ca}
Calef, D. F.; Deutch, J. M.;  {\it Annu. Rev. Phys. Chem.} {\bf
1983}, {\it 34}, 493.

\bibitem{Ri}
 Rice, S. A. {\it Diffusion-limited reactions}; Elsevier: Amsterdam,
 The Netherlands, 1985.

\bibitem{Yak}
Yakovlev, B. S.; Lukin, L. V. {\it Adv. Chem. Phys.} {\bf 1985},
{\it 60}, 99.

\bibitem{Mat}
Mataga, N.; Miyasaka, H. {\it Adv. Chem. Phys.} {\bf 1999}, {\it
107}, 431.

\bibitem{Noo}
Hong, K. M.; Noolandi, J.  {\it J. Chem. Phys.} {\bf 1978}, {\it
68}, 5163; {\it J. Chem. Phys.}  {\bf 1978}, {\it 68}, 5172.

\bibitem{Noo1}
Hong, K. M.; Noolandi, J.  {\it J. Chem. Phys.} {\bf 1978}, {\it
69}, 5026.

\bibitem{Street1}
Hong, K. M.; Noolandi, J.; Street, R. A. {\it Phys. Rev. B} {\bf
1981}, {\it 23}, 2967.

\bibitem{Que}  Que, W.;  Rowlands, J. A. {\it Phys. Rev. B}
{\bf 1995}, {\it 51}, 10500.

\bibitem{Woj}  Wojcik, M.;  Tachiya, M. {\it Radiat. Phys. Chem.}
{\bf 2005} {\it 74}, 132.

\bibitem{Mo}  Mozumder, A. {\it J. Chem. Phys.} {\bf 1968}, {\it 48}, 1659.

\bibitem{Chen}
Chen X. Y.;  Bradforth, S. E. {\it Ann. Rew. Phys. Chem.} {\bf
2008}, {\it 59}, 203.

\bibitem{Bart1}
Barthel, E. R.;  Martini, I. B.; Schwartz, J. {\it J. Chem. Phys.}
{\bf 2000}, {\it 112}, 9433.

\bibitem{Bart2}
Barthel, E. R.;  Martini, I. B.; Schwartz, J. {\it Science} {\bf
293}, 462 (2001).

\bibitem{Shkrob} Sckrob, I. A. {\it Chem. Phys. Lett.} {\bf 2004},
{\it 395}, 264.

\bibitem{Crow}
Crowell, R. A.; Lian,  R.; Shkrob,  I. A.; Bartels, D. M.; Chen, X.
Y.; Bradforth, S. E. {\it J. Chem. Phys.} {\bf 2004}, {\it 120},
11712.




\bibitem{Zhou1} Zhou, J.; Findley, B. R.; Francis, T. M.;  Nytko, E.
A.;  Braun, C. L. {\it Chem. Phys. Lett.} {\bf 2002}, {\it 362}, 63.

\bibitem{Zhou2} Zhou, J.; Shah,  R. P.; Findley, B. R.; Braun, C. L.
{\it J. Phys. Chem. A} {\bf 2002}, {\it 106}, 12.

\bibitem{Zhou3}
Zhong, C.; Zhou, J.; Braun, C. L. {\it J. Phys. Chem. A} {\bf 2004},
{\it 108}, 6842.

\bibitem{Pat} Patey, G. N.; Valleau, J. P.
{\it J. Chem. Phys.} {\bf 1975}, {\it 63}, 2334.

\bibitem{Nin} D. C. J. Chan, D. J. Mitchell, and B. W. Ninham,
{\it J. Chem. Phys.} {\bf 1979}, {\it 70}, 2946.

\bibitem{shu1}
Shushin, A. I. {\it Chem. Phys. Lett.} {\bf 1985}, {\it 118}, 197.

\bibitem{shu2}
Shushin, A. I. {\it J. Chem. Phys.} {\bf 1991}, {\it 95}, 3657.

\bibitem{shu3}
Shushin, A. I. {\it J. Chem. Phys.} {\bf 1992}, {\it 97}, 1954.

\bibitem{shu4}
Shushin, A. I. {\it Phys. Rev. E} {\bf 2000}, {\it 62}, 4688.

\bibitem{Lukin}
Lukin, L. V.  {\it J. Photochem. Photobiol. A: Chem.} {\bf 2007},
{\it 189}, 177.

\bibitem{Ashkin}
Ashkin, A.; Dziedzic, J. V.; Bjorkholm, J. E.; Chu, S. {\it Optics
Letters} {\bf 1986}, {\it 19}, 288.

\bibitem{Grier}
Grier, D. G. {\it  Nature} {\bf 2003}, {\it 424}, 810.

\bibitem{Lan}
Landau, L. D.; Lifshitz,  E. M. {\it Quantum Mechanics}; Pergamon:
Oxford, 1965.

\bibitem{shu5}
Shushin, A. I. {\it Chem. Phys.} {\bf 1990}, {\it 144}, 201; {\it J.
Chem. Phys.} {\bf 2002}, {\it 116}, 9792.

\bibitem{Abr}
Abramowitz, M.; Stigan, I. A. {\it Hanbook of Mathematical
Functions}; National Bureau of Standarts: Washington, DC, 1964.

\bibitem{Via}
Viana, N. B.; Rocha M. S.; Mesquita O. N.; Mazolli, A.; Maia Neto,
P. A.; Nussenzvieg, H. M. {\it Phys. Rev. E} {\bf 2007}, {\it 75},
021914.

\bibitem{Grier1}
Roichmann, Y.; Sun, B.; Stolarski, A.; Grier, D. G. {\it Phys. Rev.
Lett.} {\bf 2008}, {\it 101}, 128301.

\bibitem{Grier2}
Roichmann, Y.; Wong, V.; Grier, D. G. {\it Phys. Rev. E} {\bf 2007},
{\it 75}, 011407.

\bibitem{Evst}
Evstigneev, M.; Zvyagolskaya, O.; Bleil, S.; Eichhorn, R.;
Bechinger, C.; Reimann, P. {\it Phys. Rev. E} {\bf 2008}, {\it 77},
041107.


\end{thebibliography}
\end{document}